**Domain wall architecture in tetragonal ferroelectric thin films**
*Gabriele De Luca[1], Marta D. Rossell[2], Jakob Schaab[1], Nathalie Viart[3], Manfred Fiebig[1] and Morgan Trassin[1,\*]*

[1]Department of Materials, ETH Zürich, Vladimir-Prelog-Weg 4, 8093 Zurich, Switzerland
[2]Electron Microscopy Center, Empa, Swiss Federal Laboratories for Materials Science and Technology, CH-8600 Dübendorf, Switzerland
[3]Institut de Physique et Chimie des Matériaux de Strasbourg– CNRS UMR 7504, 67034 Strasbourg Cedex 2, France

[\*]E-mail: morgan.trassin@mat.ethz.ch

The concept of functional oxide interfaces was fundamentally expanded by including domain walls, interfaces separating regions with a different orientation of the magnetic or electric order within a material. [1] In contrast to grown interfaces, domain walls remain mobile after growth and external fields nucleate and annihilate them at will. Magnetic domain-wall control by spin torque transfer [2] and in race-track memories [3] is thus discussed for new types of magnetic memory. Ferroelectric domain walls, on the other hand, are explored for their locally different conductance and field effects. [4] Despite the extensive literature dealing with polar domain walls in ferroelectric crystals [5] and the recent revival of this research activity following the seminal work on conducting domain walls in multiferroic $BiFeO_3$ thin films, [6] many aspects of the physical properties of domain walls in epitaxial ferroelectric thin films are still not understood. A methodical reason is the difficulty to access distributions and morphologies of polar domains and domain walls throughout the film thickness — in thick films, buried domains and tilted domain-wall configurations are difficult to detect by scanning probe microscopy techniques. In addition, strain, oxygen vacancies and electrostatic environments are factors complicating the architecture of ferroelectric in comparison to magnetic domains walls. [7]

The variety of the domain patterns in thin films of tetragonal lead zirconate titanate, [8] the technologically most important ferroelectric, [9] illustrates this challenge. [10] Depending on the epitaxial strain state, tetragonal ferroelastic-ferroelectric thin films can either exhibit a pure out-of-plane polarized state (c-domains) or an admixture of in-plane-polarized domains (a-domains). [11] For the technological merit, the distribution of a- and c-domains is of utmost importance. [12] The usefulness of c-domains roots in their controllability by external electric fields. [13] In addition, the 180° domain walls separating them can exhibit enhanced local conductance; the origin of this is still under discussion. [14,15] In contrast, despite their critical role in the formation of (technologically promising) flux-closure domain patterns, [16] a-domains are to be avoided when 180° domain wall motion is aimed for. The 90° domain walls resulting from a-domains originating at surface dislocations are pinned and their immobility obstructs the controlled migration of 180° domain walls. [17] Even if a-domains are wanted, the controlled formation of these domains and of their associated 90° a/c domain walls in thin films requires post growth annealing [18] or substrate termination control. [19]

Hence, for devices based on domain-wall motion in tetragonal ferroelectrics, understanding and controlling the distribution and morphologies of a- and c-domains is essential. [12] First and foremost, this requires their observation. Scanning transmission electron microscopy (STEM) provides this access but it is a destructive technique and working on a length scale much smaller than the typical lateral size of domain structures. The resulting lack of quantitative information on domain patterns and domain wall morphologies impedes identification of a sound relation between structure and properties, thus obstructing nano-technological design.

Here we use a combination of STEM and optical second harmonic generation (SHG) to determine the relation between strain, film thickness, local electric fields and the resulting domain and domain-wall structure across the entire thickness of a set of PbZr$_{0.2}$Ti$_{0.8}$O$_3$ (PZT) films. We quantify the distribution of a-domains in the c-domain matrix of the films. Using locally applied electric fields we control the a/c distribution and induce the technologically preferable 180° domain walls. We find that these voltage-induced walls are tilted and exhibit a mixed Ising-Néel-type transverse rotation of polarization across the wall [7] with a specific nonlinear optical response.

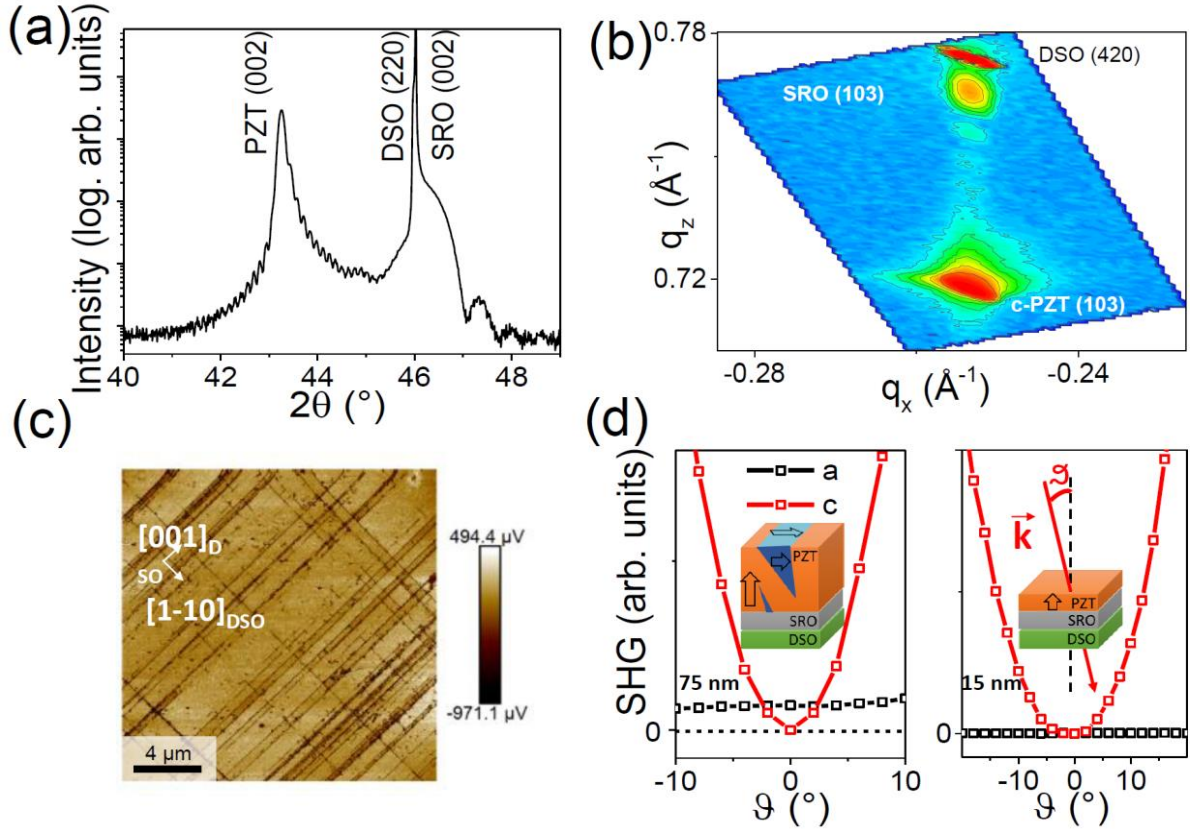

**Figure 1**

The PZT thin films were grown by pulsed laser deposition on (110)-oriented DyScO$_3$ (DSO) with an intermediate conducting (001)-oriented SrRuO$_3$ (SRO) buffer layer of 15 nm. The PZT films were grown with thicknesses above and below the threshold of ~50 nm for purely c-oriented growth. [11] The crystalline quality and strain of the tetragonal films (a, b in-plane, c out-of-plane) were characterized by x-ray diffraction. **Figure 1a** confirms epitaxial growth of a c-oriented 75 nm PZT film on SRO/DSO. Analysis of the (103) reciprocal space map in **Fig. 1b** identifies a fully in-plane-strained state, pointing to a pure c-domain configuration. The piezoresponse force microscopy (PFM) scan in **Fig. 1c**, however, reveals the presence of in-plane oriented a-domains as $[11\bar{0}]_{DSO}$-polarized stripes along $[001]_{DSO}$. Such stripes are known to be caused by the orthorhombic nature of the substrate, [18,19] but because of their low volume fraction, x-ray diffraction did not detect them here.

In order to further verify the a/c domain distribution of the films, we used SHG, i.e., doubling of the frequency of a light wave in a material. SHG is sensitive to the breaking of inversion symmetry and hence ideal for probing ferroelectric order. [20,21,22] In particular, it allows to distinguish a- and c-domains via the direction and polarization of the SHG light. [20] The SHG signal probing the c-domains (c-SHG) increases with the angle of incidence with no signal for normal incidence, $\vartheta = 0°$. Probing the a-domains (a-SHG) reveals a signal even at $\vartheta = 0°$ (see Methods).

**Figure 1d** shows the $\vartheta$-dependence of the SHG intensity for PZT films of 75 and 15 nm. For the 75-nm film we observe the characteristic $\vartheta$-dependence of c-SHG but we also detect a-SHG identifying a-domains with preferential $[11\bar{0}]_{DSO}$ polarization. For verification, the 15-nm film, which is expected to exhibit only c-domains, shows c-SHG but no a-SHG. We can use the relative intensities of a-SHG and c-SHG for quantifying the volume fraction of a-domains in the c-domain matrix (see Methods). For the 75-nm film, we find an a-domain contingent of 0.050 which is only half the value obtained from PFM (within our detection sensitivity of ±0.005). This discrepancy is a clear indication that a-domains accumulate at the surface of the PZT film where they nucleate and are readily detected by PFM. In contrast, SHG probes the entire volume of the thin film in a homogeneous way and gives a true account of the domain distribution.

We then applied an electric field to the 75 nm film via the SPM tip in order to move and induce 180° c-domains. The as-grown region of the SHG image in **Figure 2a** reveals $[11\bar{0}]_{DSO}$-polarized a-domain stripes along $[001]_{DSO}$. According to the SHG data in Figure 2a and **Figure 2b** these stripes survive in the ±10V-poled box-in-box pattern. PFM, however, reveals a single-c-domain state in the poled areas in the present case. This points to buried residual nanosized a-domains in our 75 nm film, which are only traceable with SHG. This is indeed confirmed by the high-resolution cross-section high-angle annular dark-field (HAADF) STEM image and the rotation map in the **Figures 2c and 2d**. The buried a-domains originate from the interface with the SRO where they lead to 90° domain walls. [17]

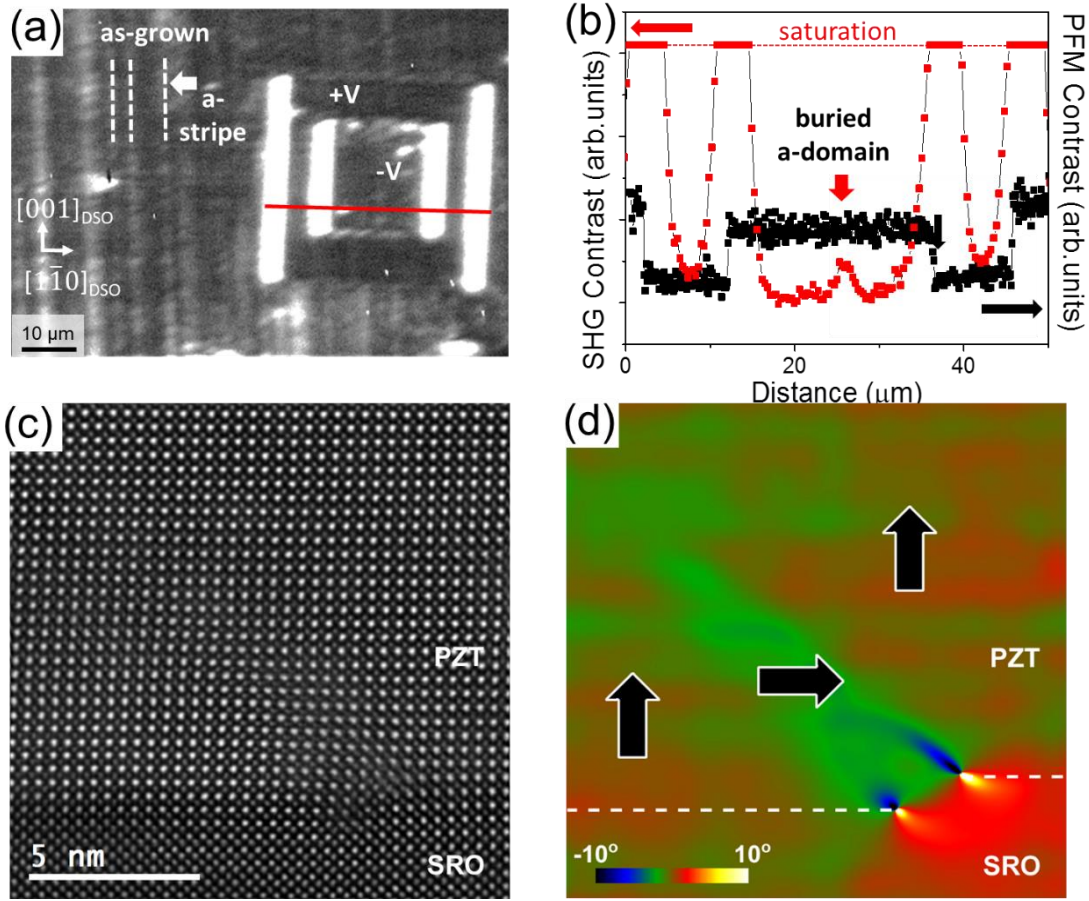

**Figures 2**

The most striking feature of Fig. 2a is the drastic SHG enhancement at the $[11\bar{0}]_{DSO}$-oriented edges of the tip-poled region. Normally, the SHG intensity at Ising-like 180° domain walls drops to zero because of an interference effect: The SHG waves from opposite domains exhibit a 180° phase difference which leads to cancelation of these contributions when they are superimposing in the vicinity of the walls. [22,23] **Figures 3a to 3c** show that the abnormal enhancement in the present study occurs at all the voltage-induced 180° c-domain walls of the tip-poled regions, yet with differently polarized SHG light from $[11\bar{0}]_{DSO}$- and $[001]_{DSO}$- oriented walls. SHG intensity enhancement induced by a reduction of symmetry has been reported at thermotropic phase boundaries in ferroelectric crystals. [24] In our case, however, the SHG selection rules associate this signal uniquely to an in-plane polarization component between the oppositely out-of-plane-polarized c-domains. This in-plane component is always polarized perpendicular to the domain walls. As discussed below, the in-plane polarization component in Figs. 3a to 3c is not caused by a-domain surface nucleation at the 180° c-domain walls. This leaves a remarkable Néel-like rotation of polarization throughout the 180° c-domain walls, recently identified to occur in PZT single crystals [25], as likely explanation for the SHG enhancement. Any Bloch-like polarization component can be indeed ruled out due to the absence of the SHG response expected for polar component along the domain wall direction (see Figure 3b and 3c). According to the large magnitude of the SHG yield, this wall extends across the entire thickness of the film (see **Figure 3e**).

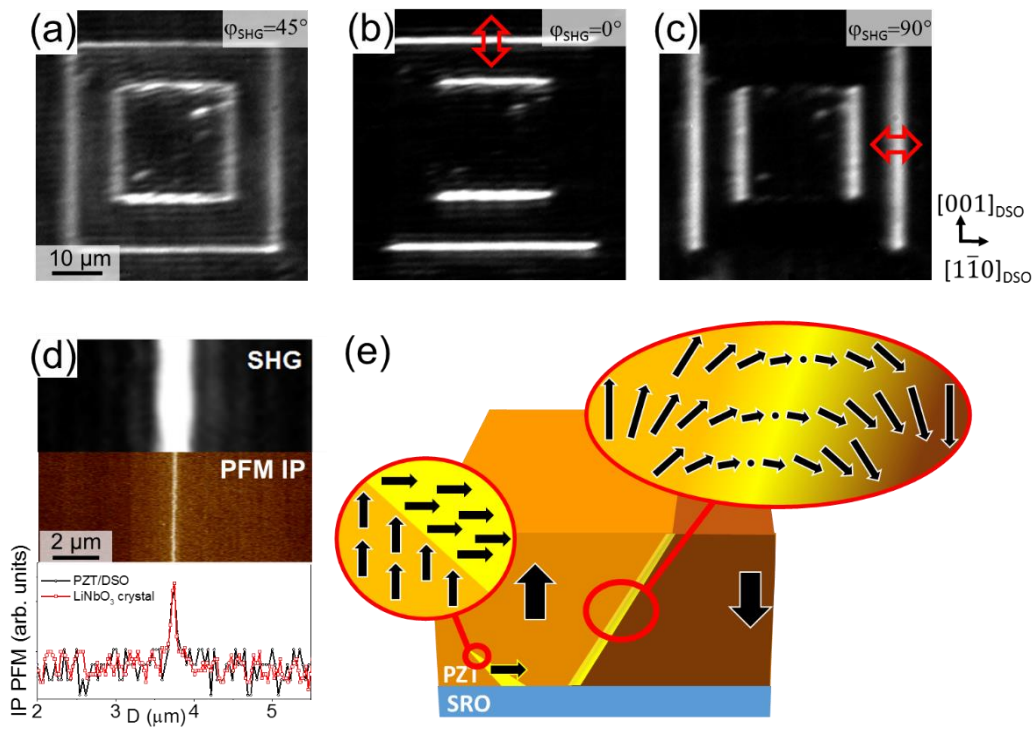

**Figure 3**

The SHG images in Figs. 2 and 3 are top views of the PZT film, probing the polar structure of the domain wall in projection onto the film surface. For obtaining a side view and the polar structure perpendicular to the film surface, we used HAADF-STEM. The HAADF-STEM image and the corresponding strain map in **Figures 4a and 4b** reveal the fine structure of the voltage-induced 180° c-domain wall.

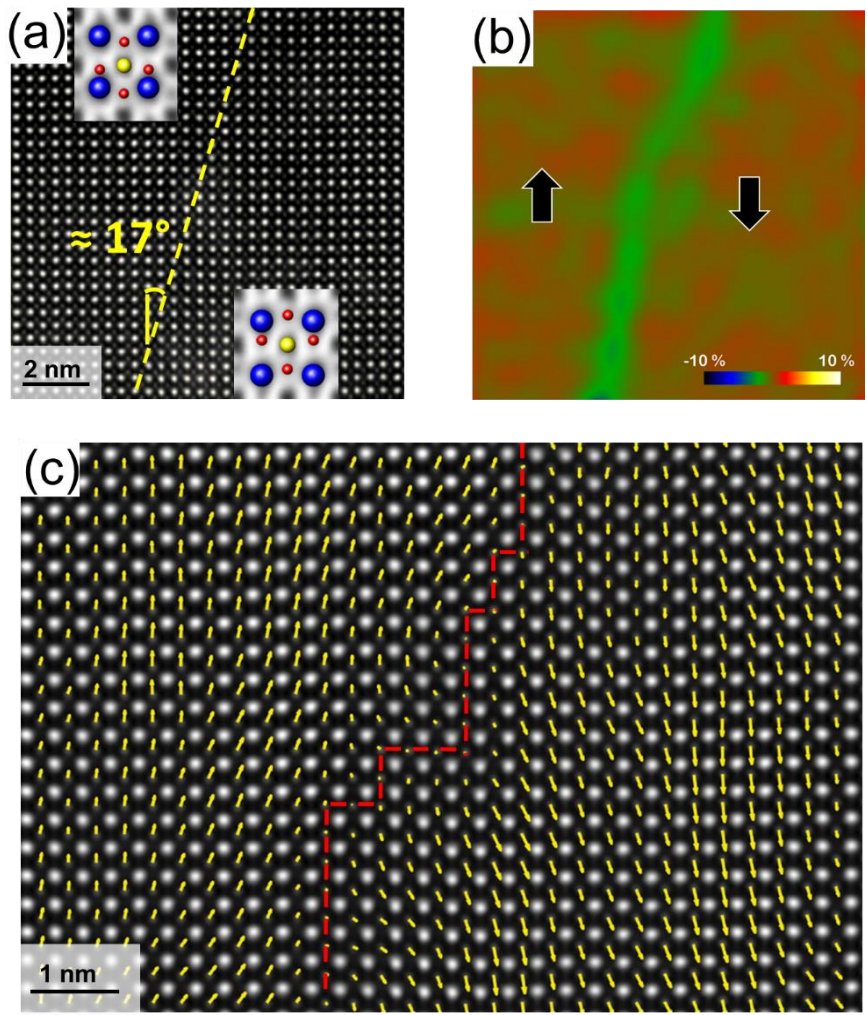

**Figure 4**

Strikingly, the tip-induced remnant wall is inclined by about 17° whereas ferroelectric PZT walls in the remnant state walls are usually running perpendicular to the film surface. [26] High-resolution polarization mapping in **Figure 4c** confirms a Néel-like reversal of the polarization across about 10 unit cells. Closer inspection also reveals an additional Ising-like variation of the polarization magnitude across the wall so that, in summary, 180° c-domain walls in the 75 nm PZT film display a mixed Ising-Néel type character. Such flexoelectric-like behavior [27] is very much unlike the sharp Ising-like 180° walls typically met in ferroelectrics. Polarization rotation of a similar, yet not identical type has been predicted for uniaxial ferroelectrics; [7] here we observe it in a PZT thin film exceeding a certain thickness. (Note that our fully c-oriented 15 nm PZT film does not show the SHG enhancement that would identify a rotation with a Néel-like component in the wall.)

We emphasize that the domain-wall structure revealed by Figs. 2 to 4 cannot be interpreted in terms of an interstitial a-domain separating oppositely polarized c-domains. Three reasons account for this. First, the PFM scan in **Figure 3d** across the region of enhanced SHG intensity in our PZT film looks identical to a PFM scan across 180° c-domain walls in a commercially available periodically poled $LiNbO_3$ crystal. This similarity between PZT and a sample in which a-domains cannot occur in the first place (the $LiNbO_3$ reference crystal) shows that, within the sensitivity of a PFM experiment, there is no interstitial a-domain present at the 180° c-domain wall. This conclusion is independent of the orientation of the walls – perpendicular to the surface or tilted. [28] Second, in Fig. 4 we do not observe the wedged end of an interstitial a-domain, as it is, for example, shown in Ref. [29]. Instead we see a sheet with a constant width of ~10 unit cells that is extended across a large distance. Third, in the center of this sheet, the polarization is zero, something that would not occur in an interstitial a-domain.

Let us finally comment on the remnant inclination of the domain wall resolved in Fig. 4. The dashed red line highlights that the wall is segmented into sections of a few unit cells that are running either parallel or perpendicular to the direction of polarization. [30] In the latter case, we obtain a tail-to-tail meeting of electric dipole moments along the domain wall. This, combined with the Ising-Néel-like rotation of polarization across the domain wall, can lead to a local accumulation of screening charges at the wall. In addition to defect-induced mechanisms [14], we therefore have to consider the three-dimensionally expanded structure of the domain walls when we discuss their conductance: their tilt with respect to the direction of polarization and the reorientation of polarization across the wall. In our tetragonal films we therefore encounter a situation that is of a similar complexity as in some bulk ferroelectrics. [25,28]

In summary, our work leads to a substantially revised view of the distribution and properties of a- and c-domains and their walls in tetragonal ferroelectric thin films using PZT, their technologically most relevant representative, as our model compound. Its a-domains are inhomogeneously distributed with accumulation at the top and bottom of the film. Specifically, poled films in an allegedly pure c-domain state may still exhibit buried a-domains. The 180° c-domain walls in these films exhibit an unusual nonlinear optical signature corresponding to a mixed Ising-Néel type rotation of polarization across about 10 unit cells and an inclination away from the film normal. The domain wall tilt leads to a macroscopic tail-to-tail polarization component. Hence, our combination of SHG with scanning probe microscopy and STEM reveals an important complementary effect to localized defects (like oxygen vacancies) that determines the properties of c-domain walls in tetragonal ferroelectrics. This is the three-dimensionally expanded structure of the domain walls: their tilt with respect to the direction of polarization and the presence of a-oriented polarization components within the wall. In ferroelectrics of technological interest, the overarching understanding of domain walls and their control thus acquired, opens new avenues towards the investigation of the dynamics of such nano-scale conduction elements and their future use in nanoelectronic device building blocks.

**Experimental Section**

*Sample preparation.* PZT Thin films were grown by pulsed laser deposition. The $DyScO_3$ substrates (Crystec GmbH) were kept at 700°C during the $SrRuO_3$ deposition and 600°C for the PZT growth. The oxygen partial pressure was constant at 0.1 mbar. The KrF excimer laser fluence was set to 0.9 J/cm$^2$ and the laser repetition rate was 8 Hz.

*Structural characterization.* The film orientation, crystal quality, strain state and thickness were studied by x-ray diffraction and reflectometry using a Smart lab Rigaku diffractometer equipped with a rotating Cu anode source operating at 9 kW, parallel-beam optics and a four-bounce Ge 220 monochromator producing Cu-K$\alpha$1 radiation at 1.5406 Å.

STEM was carried out using a double spherical aberration-corrected Jeol JEM-ARM200F microscope operated at 200 kV. The convergence semi-angle was set to 25 mrad. In typical operating conditions for the experiments described in this paper, the microscope provides an estimated spatial resolution of 0.8 Å. The annular semi-detection range of the annular dark-field detector was set to collect electrons scattered between 90 and 370 mrad for the HAADF images and between 14 and 27 mrad for the annular-bright-field images. All high-resolution STEM images used for polarization mapping were filtered in the Fourier space using a Gaussian grid mask on selected reciprocal lattice frequencies. The positions of the atomic columns were directly determined on the image by using the Peak Pair Analysis software package. Subsequently, the polar displacements in the image plane of the Zr/Ti cations were measured relative to the center of the surrounding Pb cations using a home-developed code running in MatLab MATLAB R2014b. The polarization vectors in the polarization maps are plotted opposite to the polar displacement of the Zr/Ti cations which point toward the center of the negative oxygen charges. Geometrical phase analysis was performed by using the GPA software package for Digital Micrograph. The sample for the STEM analysis was prepared across the +10/-10 V poled 180° domain wall by means of a FEI Helios Nanolab 450S focused ion beam.

*Second harmonic generation.* SHG is a nonlinear optical process denoting the emission of light at frequency $2\omega$ from a crystal irradiated with light at frequency $\omega$. This is expressed by the equation $P_i(2\omega) = \varepsilon_0 \Sigma_{j,k} \chi^{(2)}_{ijk} E_j(\omega) E_k(\omega)$, where $E_{j,k}(\omega)$ and $P_i(2\omega)$ are the electric-field components of the incident light and of the nonlinear polarization, respectively, with the latter acting as the source of the SHG wave. The nonlinear susceptibility $\chi^{(2)}_{ijk}$ characterizes the ferroelectric state. Ferroelectrics like PZT have the point-group symmetry *4mm*. For a spontaneous polarization along the z axis, the allowed SHG tensor components are $\chi^{(2)}_{zzz}$, $\chi^{(2)}_{zxx} = \chi^{(2)}_{zyy}$, $\chi^{(2)}_{xxz} = \chi^{(2)}_{xzx} = \chi^{(2)}_{yyz} = \chi^{(2)}_{yzy}$. For the c-domains the association is a, b, c → x, y, z. For a-domains with the spontaneous polarization along the a- or b-axis the associations are a, b, c → z, x, y and a, b, c → y, z, x, respectively. We see that at least one of the light waves contributing to SHG needs to be polarized along the direction of the ferroelectric polarization. Therefore c-domains do not yield SHG in normal incidence and all the SHG signal in this configuration is related to a-domains. In non-normal incidence, the c-domain-related SHG contribution will increase with increasing tilt angle $\vartheta$. An appropriate set of measurements performed on a single-c-domain sample with different polarizations of the light waves at ω and 2ω and different angles $\vartheta$ of the incident light therefore revealed the values of all the tensor components $\chi^{(2)}_{ijk}$ at our probe wavelength. Once we know these tensor components, measurements on multi-domain samples will not only allow us to separate the SHG contributions emitted from a- and c-domains but, furthermore, to determine the volume ratio of the different domain states from the relative amplitude of a- and c-domain-related SHG yields. For probing the PZT films, we used light pulses emitted at 1 kHz from an amplified Ti:sapphire system with an optical parametric amplifier. The light pulses had a photon energy of 0.95 eV, a pulse length of 120 fs and a pulse energy of 20 μJ. The setup for SHG is described in Refs. 21 and 22.

**Acknowledgements**

The authors thank D. Meier and Th. Lottermoser for fruitful discussions and S. Manz for experimental assistance. The authors acknowledge funding through the SNSF R'Equip Program (No. 206021-144988). Access to the TEM facilities at the IBM Research-Zurich, Switzerland, under the IBM/Empa Master Joint Development Agreement is gratefully acknowledged.

**Figure captions**

**Figure 1** . **a,** X-ray diffraction measurement in the θ-2θ configuration. **b,** The reciprocal space map around the (103) reflection shows that the PZT films are coherently strained. **c,** An out-of-plane PFM scan reveals stripe-like a-domains on a c-domain background. **d,** SHG yield as a function of the tilt angle $\vartheta$ for PZT films of 75 nm and 15nm. The thicker film reveals c-SHG and a-SHG, the thinner film only c-SHG. Insets depict sketches of the samples with their respective ferroelastic domain configuration.

**Figure 2** a SHG image showing $[1\bar{1}0]_{DSO}$-polarized a-domain stripes oriented along the $[001]_{DSO}$ direction in the as-grown region of the 75-nm PZT film. A voltage-induced box-in-box pattern is also visible. **b,** Line scans by PFM (black) and SHG (red) of the cross section shown as red line in (a). The SHG scan reveals a buried a-domain in the poled regions that is invisible to PFM. **c,** HAADF-STEM image confirming the presence of a buried a-domain in the poled area. The a-domain emerges from two dislocations at the PZT/SRO interface. **d,** Rotation map of the image in (c) obtained by geometric phase analysis. Block arrows indicate the rotation of the polarization at the needle-like a-domain.

**Figure 3 a,** An image taken with SHG light polarized under $\varphi_{SHG} = 45°$ (with $\varphi_{SHG} = 0°$ as a-axis) shows all the voltage-induced 180° c-domain walls of the 75-nm PZT film. **b,** Selective detection of c-domain walls with a $[001]_{DSO}$ polarization component (arrow) with SHG light at $\varphi_{SHG} = 0°$. **c,** Selective detection of c-domain walls with a $[1\bar{1}0]_{DSO}$ polarization component (arrow) with SHG light at $\varphi_{SHG} = 90°$. **d,** Comparison between SHG intensity (upper panel) and in-plane PFM response (lower panel) at a 180° domain wall. The line scan compares the in-plane PFM response of a 180° domain wall of a PZT film and a LiNbO$_3$ crystal. **e,** Schematic of a buried a-domain and a ferroelectric mixed Ising-Néel-type domain wall.

**Figure 4 a,** HAADF-STEM image of a 180° c-domain wall which reveals to be inclined by 17°. The insets are annular bright-field STEM images with overlaid structural models showing the displacement of the oxygen (red) and Zr/Ti (yellow) atomic columns from the center of the surrounding Pb (blue) cations. **b,** Corresponding strain color map (strain tensor $\varepsilon_{xy}$) determined by geometric phase analysis of the domain wall in (a). The arrows show the direction of polarization. **c,** HAADF-STEM image overlaid with polarization vectors showing a smooth Néel-like rotation combined with an Ising-like polarization decrease across the domain wall. The dashed line indicates the polarization minimum and hence the center of the domain wall.